\pdfoutput=1
\PassOptionsToPackage{bookmarksnumbered,unicode}{hyperref}
\RequirePackage{hyperref}
\documentclass[sigconf]{acmart}
\usepackage{tabularx}
\usepackage{makecell}

\AtBeginDocument{%
  \providecommand\BibTeX{{%
    \normalfont B\kern-0.5em{\scshape i\kern-0.25em b}\kern-0.8em\TeX}}}

\copyrightyear{2024}
\acmYear{2024}
\setcopyright{rightsretained}
\acmConference[CHI '24]{Proceedings of the CHI Conference on Human Factors in Computing Systems}{May 11--16, 2024}{Honolulu, HI, USA}
\acmBooktitle{Proceedings of the CHI Conference on Human Factors in Computing Systems (CHI '24), May 11--16, 2024, Honolulu, HI, USA}
\acmDOI{10.1145/3613904.3642894}
\acmISBN{979-8-4007-0330-0/24/05}

\acmSubmissionID{7887}

\usepackage{array}
\newcolumntype{L}[1]{>{\raggedright\let\newline\\\arraybackslash\hspace{0pt}}m{#1}}
\newcolumntype{C}[1]{>{\centering\let\newline\\\arraybackslash\hspace{0pt}}m{#1}}
\newcolumntype{R}[1]{>{\raggedleft\let\newline\\\arraybackslash\hspace{0pt}}m{#1}}

\begin{document}

\title[Envisioning Large Language Model Use by Autistic Workers for Communication Assistance]{``It’s the only thing I can trust'': Envisioning Large Language Model Use by Autistic Workers for Communication Assistance}

\author{JiWoong (Joon) Jang}
\email{jwjang@cmu.edu}
\orcid{0000-0003-0469-9501}
\affiliation{%
  \institution{Human-Computer Interaction Institute\\ Carnegie Mellon University}
  \city{Pittsburgh}
  \state{PA}
  \country{USA}
}

\author{Sanika Moharana}
\email{smoharan@cs.cmu.edu}
\orcid{0009-0007-4899-5309}
\affiliation{%
  \institution{Human-Computer Interaction Institute\\ Carnegie Mellon University}
  \city{Pittsburgh}
  \state{PA}
  \country{USA}
}

\author{Patrick Carrington}
\email{pcarrington@cmu.edu}
\orcid{0000-0001-8923-0803}
\affiliation{%
  \institution{Human-Computer Interaction Institute\\ Carnegie Mellon University}
  \city{Pittsburgh}
  \state{PA}
  \country{USA}
}

\author{Andrew Begel}
\email{abegel@cmu.edu}
\orcid{0000-0002-7425-4818}
\affiliation{%
  \institution{Software and Societal Systems Department\\ Carnegie Mellon University}
  \city{Pittsburgh}
  \state{PA}
  \country{USA}
}

\renewcommand{\shortauthors}{Jang et al.}
\begin{abstract}
Autistic adults often experience stigma and discrimination at work, leading them to seek social communication support from coworkers, friends, and family despite emotional risks. Large language models (LLMs) are increasingly considered an alternative. In this work, we investigate the phenomenon of LLM use by autistic adults at work and explore opportunities and risks of LLMs as a source of social communication advice. We asked 11 autistic participants to present questions about their own workplace-related social difficulties to (1) a GPT-4-based chatbot and (2) a disguised human confederate. Our evaluation shows that participants strongly preferred LLM over confederate interactions. However, a coach specializing in supporting autistic job-seekers raised concerns that the LLM was dispensing questionable advice. We highlight how this divergence in participant and practitioner attitudes reflects existing schisms in HCI on the relative privileging of end-user wants versus normative good and propose design considerations for LLMs to center autistic experiences.
\end{abstract}

\ccsdesc[500]{Human-centered computing~Empirical studies in accessibility}

\keywords{Autism, social communication, large language models, workplace, Artificial Intelligence, Neurodiversity}



\begin{teaserfigure}
    \includegraphics[width=\textwidth]{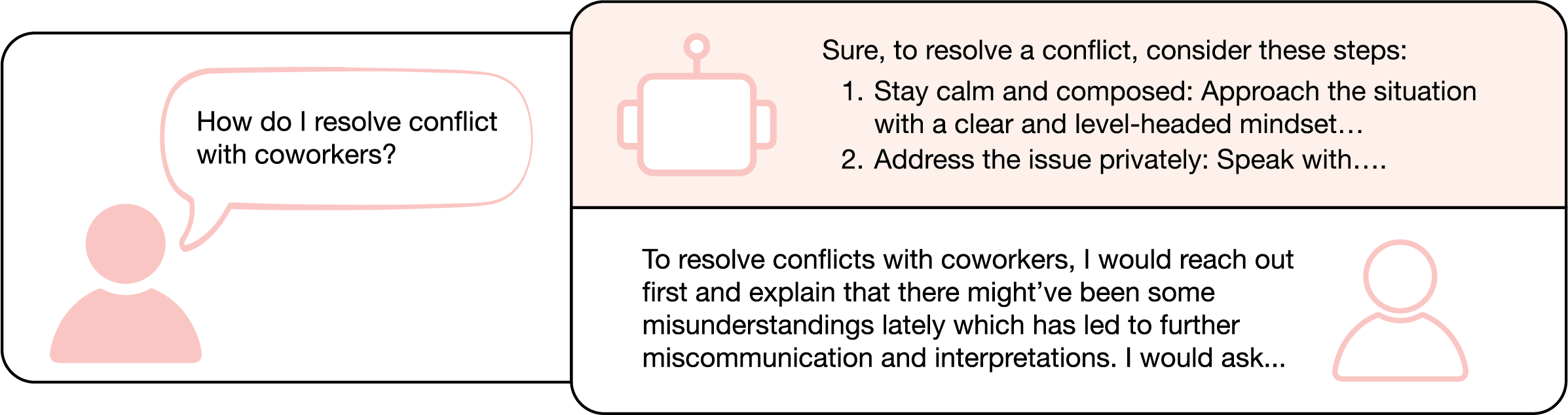}
  \caption{We investigate opportunities and risks from autistic workers' use of LLMs for assistance with social communication and scenarios and had participants juxtapose LLM and human confederate-generated responses. A question from one of our participant sessions is displayed, along with corresponding LLM and human confederate replies.}
  \Description{A graphic displaying a participant asking the question: "I'm about to go to a conference. How do I go about forming connections?", and two replies, one from a robot and one from a human are juxtaposed.}
  \label{fig:teaser}
\end{teaserfigure}
\maketitle

\section{Introduction}
\label{sec:intro}
Despite wide recognition, research, and advocacy, autistic\footnote{While some in the autism community prefer people-first language (PFL), others have embraced the term ``autistic'' as their chosen identifying label, and we have opted for that terminology as well, though we acknowledging the preference for PFL among some \cite{Kenny2015}.} adults continue to face critical levels of employment challenges~\cite{krzeminska_austin_bruyère_hedley_2019, WALKOWIAK2021}. Prior work has shown that up to 85 - 90\% autistic adults face unemployment and underemployment issues~\cite{Taylor_2010, Roux2015}, among the highest of all major disability groups in the United States~\cite{Leven2020}. When seeking employment and navigating the workplace, autistic workers report that they often face challenges such as understanding non-autistic social norms, handling limited or unclear information and expectations, discerning unspoken meaning in communication, coping with sensory overload, and managing with limited support and accommodations. Furthermore, many face difficulties navigating issues such as disclosure and advocating for roles which take advantage of their unique characteristics~\cite{Rebholz2012, Bissonnette2013-zn, Leven2020}.

Given that autistic individuals experience challenges with social interaction in the workplace~\cite{apa2013, Rebholz2012, Bissonnette2013-zn}, researchers have noted the importance of having a champion or advocate who is able to mediate and assist the autistic individual in social or communication challenges at work~\cite{Nicholas2019}. 
Some research has found positive effects that specialized job coaching for autistic job-seekers and counseling with speech and language pathologists (SLPs) can have~\cite{Scott2015, Fong2021, Nicholas2019}, but given the challenges and research highlighting the need for more affordable and better support~\cite{Rebholz2012,WALKOWIAK2021,krzeminska_austin_bruyère_hedley_2019,Scott2015}, it appears unlikely that autistic workers will access essential resources in most jobs.
This underscores the desire for a readily accessible resource that autistic workers can turn to in moments of social difficulty -- a desire which we explore whether large language models (LLMs) can support.

Our work assesses the capability of LLMs to intermediate and otherwise assist tactfully with workplace communication and related acts. This is motivated by LLMs' recent astronomic rise in popularity in adoption coupled with its apparent ability to simplify and explain social interactions. These include reports of utilization in contexts such as writing ``how-tos'' for social interactions, interpreting and explaining social situations, explaining humor in jokes, and proofreading workplace communications by the broader public~\cite{harwell_tiku_oremus_2022}. Much of this is attributable to the fact that LLMs encode a wide range of human behavior in its training data corpus~\cite{stoparrot}. More recent chatbot incarnations (e.g. OpenAI's ChatGPT~\cite{openai_2022} and Google's Gemini / Bard~\cite{pichai_2023}) have employed techniques like reinforcement learning from human feedback (RLHF)~\cite{rlhf2022} to achieve an emergent capability to generate outputs with increased perceived social awareness and coherence compared to prior language models. In autism-allied and -focused spaces in social media and online forums [e.g. Facebook, X (\textit{formerly Twitter}), TikTok, Reddit, etc.], this apparent capability has not gone unnoticed: a growing number of posts now involve users opining, speculating, and relaying their experiences after having used LLMs in social communication contexts~\cite{wiredSomeAutistic}. Many explicitly note utilizing LLM-based tools like ChatGPT to understand and prepare for social situations at work~\cite{wiredSomeAutistic, lilgreycat2023chatgpt, johnston2023chatgpt}, write emails and messages to supervisors and coworkers~\cite{johnston2023chatgpt, comprehensivetime6342023gpt}, and understanding vague communications and instructions~\cite{comprehensivetime6342023gpt, sacbh2023chatgpt} -- with one thread dubbing it ``a gamechanger for people on the spectrum~\cite{sacbh2023chatgpt}.'' Given this, it can no longer be assumed that this use case is mere hypothetical.

To better envision the opportunities and concerns with autistic users utilizing LLMs for assistance in workplace communication, we aim to answer the following research questions.

\begin{enumerate}
\label{sec:RQs}
    \item[\textbf{RQ 1}] What communication challenges and resource availabilities (or deficits) do autistic workers experience?
    \item[\textbf{RQ 2}] Do autistic workers believe that an LLM's advice could be helpful for addressing workplace communication challenges?
    \begin{enumerate}
        \item[\textbf{2a}] If so, why?
        \item[\textbf{2b}] If so, how would autistic workers utilize LLMs?
    \end{enumerate}
    \item[\textbf{RQ 3}] Can an LLM's advice be considered \textit{good}? (and in what / whose definition of
 \textit{good}?)
\end{enumerate}

We conducted a within-subjects study with (\textit{n} $=11$) autistic individuals where we encouraged participants to maintain a free dialogue with two chatbots, an LLM (\textit{utilizing OpenAI's GPT-4~\cite{gpt4} via API with some prompt engineering}) and a human confederate in disguise. During the study, participants (1) shared with us their prior experiences with workplace communication challenges and available resources, (2) engaged in exchanges with the chatbots, and (3) rated and described their preferences for and between the LLM and confederate. 

Overall, our data shows that participants showed a strong affinity to the LLM, with \textit{nine (9) out of 11 $(\simeq 82\%)$} expressing a desire to continue using it for communication aid and social advice. \textbf{We found the same nine (9) out of 11 $(\simeq 82\%)$ participants preferred LLM interactions to ones with the human confederate} through quantitative and qualitative analysis. Participants' motivations to continue using the LLM were informed by a lack of available resources at work, with many relying on friends and family despite prior experiences of emotional harm. 

Participants valued the LLM for its potential to communicate in ways which ascribed to their preferences, untangle implicit neurotypical norms, and allow for the freedom to ask questions without fear of reprisal -- providing a sense of control in navigating the neurotypical world. To provide grounding of the LLM's answers  vis-à-vis alternative resources to autistic workers, we had a professional counselor and job coach (\textit{LPC, NCC}) specializing in workforce readiness training for neurodivergent individuals evaluate LLM and confederate responses. While the practitioner placed value on the potential for LLMs' ease-of-access in times of need, she noted the LLM's tendency to make ungrounded assumptions and assume neurotypicality, and raised concerns about harms resulting from the LLM's misaligned advice and participant acceptance of potentially harmful recommendations.

From our results, we note a divergence in attitudes towards the LLM: \textbf{one where autistic participants express a desire to use an LLM because it appears to provide agency for independent exploration, while our practitioner urges caution due to its misleading advice.} This division symbolizes and foreshadows imminent societal concerns as LLMs are considered by autistic users~\cite{redditRedditDive} and are being developed explicitly for interpersonal advice~\cite{nytimesLLMLifeAdvice}. Despite its novel façade, we show how the disagreement between participants and practitioner mirrors existing conflicts in HCI literature from accessibility, disability theory, and medical/social models of disability --- ones involving the relative privileging of disabled experiences against normative authorities. We believe acknowledging this parallel allows for progress towards \textit{truly assistive} aids for autistic workers which offers access to grounded advice while centering their lived experience.

This work presents the following exploratory contributions which we believe merits further consideration:
\begin{enumerate}
    \item we illuminate the practice by autistic workers of obtaining social communication advice from LLMs (§~\ref{sec:intro}),
    \item we gauge autistic participants' preferences for receiving advice from an LLM versus a human confederate and current resources in a user study (§~\ref{sec:participant-reactions}),
    \item we evaluate the quality of LLM advice during a discussion with a specialized counselor and job coach (§~\ref{sec:practitioner-assessment}),
    \item we provide discussions of potential reasons, opportunities, and concerns for future use of LLMs for autistic workers' social communication assistance (§~\ref{sec:discussion}),
    \item and we elucidate design considerations as LLMs are considered for social communications, and demonstrate how designing LLMs for social advice is fraught with entanglements and relative privileging. (§~\ref{sec:design-considerations})
\end{enumerate}

\section{Related Works}
\subsection{Autism in the Workplace}
\label{sec:autism-in-the-workplace}
Autistic adults not only experience the highest level of unemployment amongst all major disability groups in the United States, but they also tend to be the least considered in any company diversity or disability efforts, reinforcing the large employment gap for autistic adults~\cite{harmuth2018barriers}. When seeking employment and navigating the workplace, autistic workers report that they often face challenges such as overwhelming sensory processing limitations, and accessing accommodations~\cite{WALKOWIAK2021, Roux2015}. Apart from these, they face greater invisible social difficulties including understanding non-autistic social norms, ambiguous expectations, discerning unspoken meaning in communication~\cite{Nicholas2019} that result in negative experiences~\cite{szechy2023autism, lopez2022commentary}.  These contrasting communication practices and preferences between autistic and non-autistic people pose a mismatch across social interaction as exemplified by Milton’s Double Empathy perspective~\cite{milton2012ontological}.

Research has highlighted the significance of autistic individuals having a champion or advocate in their corner to assist with workplace communication challenges~\cite{nittrouer2022career}, but they also might turn to family, friends, and peers for daily living activities and advice, or find other forms of social support networks online that help meet everyday socioemotional needs~\cite{chan2023beyond, ghanouni2023resilience}. However, these personal resources do not always possess the aptitude to provide guidance for workplace dilemmas. 

Within the workplace, the landscape of resources for autistic people in mitigating these difficulties can include job coaches, vocational rehabilitation specialists, occupational therapists or licensed professional counselors (LPC’s)~\cite{Fong2021}. These experts can provide personalized support for autistic individuals across their employment, career pathways, and professional development journeys, but are largely inaccessible and not available at scale for autistic adults~\cite{grob2019assessing, walsh2014employment}. There are also community organizations, government agencies, and nonprofits that offer employment readiness programs and support services~\cite{nicholas2018ecosystem, gerhardt2011addressing}. Many workplaces are integrating wider autism at work initiatives and efforts within human resource departments to advance assistance~\cite{johnson2020workplace}. However access and availability of workplace accommodations vary across domains, industries of work, and their organizational policies. With the number of adults with autism seeking support increasing, many are unable find resources, services, and programs through the aforementioned avenues~\cite{burgess2014employment}. 

Considering the discrimination due to communication differences that autistic individuals experience, most development of technological aids and tools are directed towards social engagement, educational goals, and learning experiences~\cite{wang2023assistive}. Many online communication platforms, internet forums, and social media have been beneficial for autistic people in cultivating connectedness, social relationships, and seeking support in a controllable manner~\cite{burke2010social, hassrick2021benefits}, especially through written and text-based mediums~\cite{howard2021anything}. Emerging technologies that have surfaced hyperfocus on autistic children and support for parents in promoting social communication via voice assistants, smart glasses, AAC devices, and other speech-generated outputs~\cite{schepis1998increasing, keshav2017social, schlosser2015speech}. With popular investment of early interventions for autistic children~\cite{zhang2018design, heng2021research}, there is an increasing amount of negligence towards autistic adults and a lack of tools for supporting their social communication needs~\cite{cummins2020autistic}.  In the context of the workplace, interactive communication technologies aimed to help adults in interviewing, the job search, and pre-emptive training process can be occasionally found,~\cite{hayes2015mobile}, but tools for longitudinal support and retention in the workplace are scarce~\cite{baldwin2014employment}.

Another critical element in finding support is dependent on disclosure. Disclosure of one's autism can be an increasingly vulnerable and stigmatizing experience even amongst immediate friends and family~\cite{togher2023disclosing, frost2019just, thompson2020helpful}. Disclosure in the workplace poses further potential risks to job security, unwarranted biases and microaggressions, and threats to acceptance~\cite{Leven2020}. Within workplaces, the situation may vary --- larger and more structured workplaces have designated HR professionals to support employees in requesting and receiving accommodations, while others may mandate speaking directly to a supervisor or manager. This establishes more power differentials and fear of transparency, placing a greater burden on autistic employees to navigate on their own. Considering these disclosure challenges~\cite{lindsay2021disclosure}, coupled with the plethora of research which speaks to the need for improved affordable support structures for addressing social difficulties, it is unlikely that many autistic workers would have access to the kinds of resources they need. 


\subsection{HCI and Autism}
\label{sec:hci-autism}
Within HCI, designing for neurodiversity and technology for supporting autistic individuals has been explored in most popular applications through web search and navigation, virtual reality, software development, visual tools, and especially computer-mediated communication tools such as social network and media platforms, video calling, texting and messaging~\cite{zolyomi2019managing, burke2010social, taylor2022autism, lagerqvist2020seen, glaser2023video}. There has been research devoted to conducting participatory design work with autistic individuals and assistive technology, identifying the challenges that inclusive and universal design methods face when designing for autistic people and ways of better collaborating with neurodivergent adults.~\cite{waardenburg2022design}. However, significant amounts of HCI research relating to autism have focused on autistic children, who have different communication patterns and needs than adults.

Within the workplace context, Annabi et al. have characterized autism employment across experiences for women, software developers and other IT professionals while creating frameworks and context through which adjustments, accommodations, and support for these autistic workers can be explored to create more inclusive learning and mitigating such barriers~\cite{annabi2019theoretical, romeijnexamining, annabi2018untold}. Reviews of accessibility research in HCI~\cite{mack-lit-review} and particularly neurodiversity research, centers around the mismatches of neurodivergent users using generalized technology or how assistive technology can drive behavior change for these, but few focus on consequences of such interventions. 

\subsection{Autism, Generative AI, and LLMs}
\label{sec:autism-LLM}
In the rapidly evolving landscape of generative AI and LLMs, HCI research contemplates the possibilities of its application towards various domains of use, along with developing guidelines~\cite{weisz2023toward} and exploring its boundaries. The implications for capability, agency, and control are questioned as the quality of generative AI outcomes begin to parallel or at times surpass human strengths~\cite{muller2022genaichi}. While some researchers imagine its contributions as a design space, material, and avenue of prototyping for human-centered computing systems~\cite{ringel2023design}, a lot of attention has been devoted to how its changing the landscape of learning and educational systems~\cite{kasneci2023chatgpt}. Others have suggested its potential for supporting communication for people with disabilities, such as enhancing experiences for AAC users~\cite{valencia2023less}, writing AI-assisted emails for adults with dyslexia~\cite{goodman2022lampost} and even speech-language pathologists in delivering therapy to autistic individuals in clinical settings~\cite{du2023generative}. 

However, researchers also caution towards malicious uses, negative impacts, and reliability of LLM-generated content over its low cost and high speed~\cite{hamalainen2023evaluating, giri2023exploring}. Its efficiency is contrasted with the ethical boundaries of human augmentation or automation~\cite{byun2023dispensing} for future applications. Gadiraju et al. provide disability-centered perspectives on LLMs through discussions with people with various disabilities to formulate more disability-positive data representation strategies for diminishing societal harms and biases perpetuated by such neural language model-based dialog systems~\cite{gadiraju2023wouldn}. Despite these modern advances, it remains the case that the risks highlighted by Bender et al.~\cite{bender2021dangers}, of exposure to encoded biases and discriminatory viewpoints, lack of grounded communication, and overreliance on automated systems likely still apply with current and future human-LLM interactions, potentially exposing autistic users to said harms~\cite{whittaker2018ai}.

\subsection{Viewpoints on Technology’s Role, Design, and Relationship with Disability}
\label{sec:critical-hci-autism}
Because the subject matter of this work involves a technology's role in managing a disabled person's social representation, we introduce viewpoints which speak to worldviews on disability, and subsequently, technology's role, particularly when its harms are taken into account. We do so to acknowledge the discourse around changing definitions and interpretations of disability, which have prompted many to reassess how research includes or excludes people with disabilities’ representation and how as it maps onto technology design and development. 

\textit{The Medical Model of disability}, as the most mainstream perspective, views disability as something an individual tackles for themselves, and generally advocates for utilizing existing medicalized structures, such as therapy, or counseling~\cite{brisenden1986independent}. From this perspective, some help, despite being coupled with a loss of individual agency, may be seen as better than no help at all, especially if the benefits to said technology aligns with the model's focus on management and mitigation of disability-related challenges ~\cite{marks1997models}. The model might also lean towards the goal of assistive technology being designed for an organizational or societal good, as it often promotes strategies that help individuals conform to existing structures and norms. 

In contrast, the \textit{Social Model of disability} sees disability as a societal issue caused by barriers and attitudes within society and in general would reject the premise of relying on technological help for human-human social assistance --- arguing for a societal or organizational-encompassing solution to address unmet needs of disabled individuals~\cite{shakespeare2006social, oliver2013social}.

Relatedly, \textit{Standpoint Theory}, advocated by scholars like Lucy Suchman and Shaowen Bardzell and originating from feminist theory, remains influential in current thinking on technology design processes. It highlights the significance of individual perspectives and experiences, emphasizing that the primary objective of assistive technology is to advance the interests of the individual~\cite{suchman2016organizing, bardzell2011towards}.

\textit{Disability Studies and Critical Disability Studies}, drawing from works like those of Mankoff et al., Hamraie, and Williams et al., go further, with the goal for technology to challenge existing power structures and narratives about disability and advocate for amplifying and enabling the voices and experiences of disabled individuals~\cite{williams2023counterventions, mankoff-dis-studies, hamraie2019crip}. These perspectives serve as ones which most directly challenge the medical model of disability's view on technology as being permissive or enforcing the status quo. 

Each of these perspectives relate to differing goals and resulting tradeoffs they posit that assistive technology ought to exhibit. We explore the implications of these perspectives as we investigate autistic workers' potential use for LLMs in social and communication assistance.

\section{Methodology}
In the study, autistic workers were asked to share prior workplace communication challenges and resources, and to interact with and rate two chatbots: (1) \textit{Paprika}, a chatbot which utilized OpenAI's GPT-4~\cite{gpt4} API (\textit{with some modifications described in  §~\ref{sec:Paprika}}) and (2) \textit{Pepper}, a human confederate. A study design with a human confederate was chosen as a comparative baseline for evaluating the LLM as it represents a close analog for a readily available resource that provides human-like advice, akin to an anonymous help hotline (in chat form), Reddit comments, the social network Blind~\cite{Blind}, as well as interaction behavior which were described in social media posts and threads. To ensure participants would feel comfortable providing candid feedback without worrying about potentially offending a real person, we communicated that both \textit{Paprika} and \textit{Pepper} were automated agents. After the conclusion of the study period, we consulted with a professional counselor and job coach (\textit{LPC, NCC}) specializing in workforce readiness training for neurodivergent individuals for their opinions to reveal additional context into the efficacy and safety of LLM-generated advice.

\begin{figure*}[ht]
    \centering
    \includegraphics[width=\textwidth]{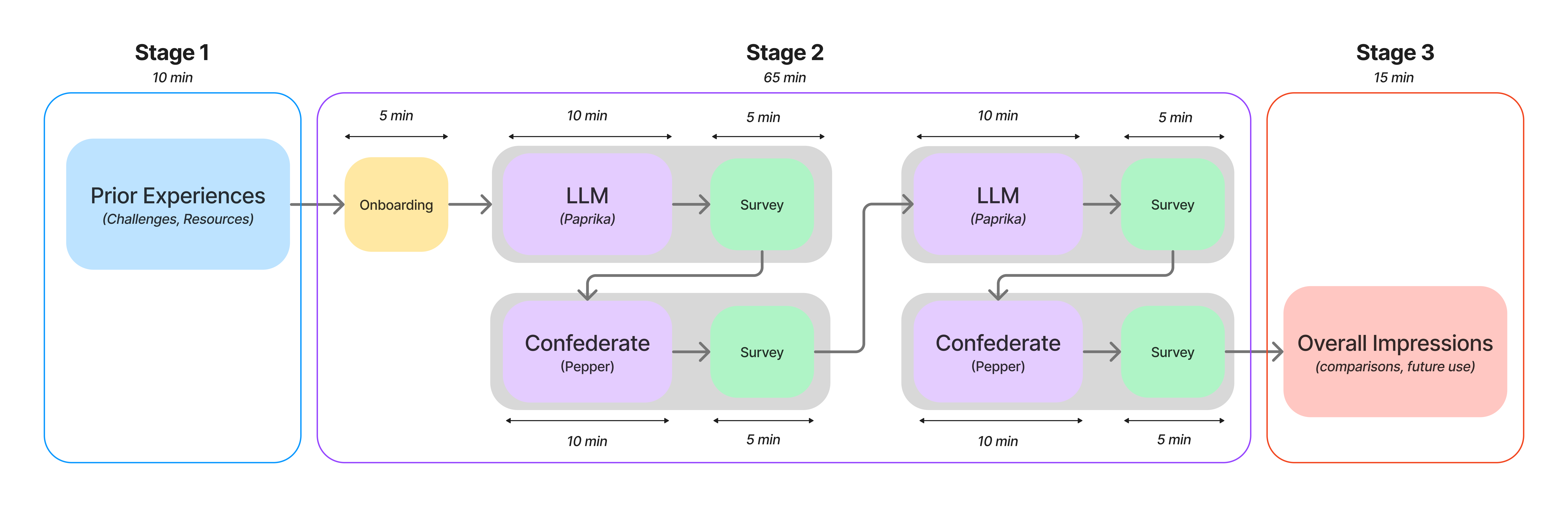}
    \caption{A diagram outlining the study protocol used. The intervention ordering was counterbalanced between participants and participants engaged with a different chatbot each session.}
    \Description{A diagram outlining the study protocol. There are three boxes indicating three phrases. The first contains a box titled Prior Experiences. The second, the intervention stage, has an onboarding box, then 4 copies of an intervention like LLM or Confederate followed by a survey. Finally, the third box contains the final stage, titled Overall Impressions.}
    \label{fig:study-flow}
\end{figure*}

\subsection{Study Protocol}
The study, whose design was approved by an Institutional Review Board (IRB),  was conducted remotely, and was designed to take 90 minutes across three parts for each participant. Data was recorded via video conference software recording, and collected video and audio data from participants, as was text inputted to the Discord space used for chatbot interactions and surveys (§~\ref{sec:data-collection}). Each participant encountered the following three stages:
\begin{enumerate}
    \item[\textbf{(1)}] a 10-minute semi-structured interview (§~\ref{sec:initial-interview}),
    \item[\textbf{(2)}] a 65-minute session interacting with and evaluating two chatbots (§~\ref{sec:chatbot-intervention}) which included,
    \begin{enumerate}
        \item[\textit{a})] a 5-minute onboarding to familiarize the participant with Discord and interacting with \textit{Paprika} (LLM) and \textit{Pepper} (human confederate),
        \item[\textit{b})] four (4) sessions of an (up to) 10-minute interactions with a chatbot and a rating survey,
    \end{enumerate}
    \item[\textbf{(3)}] and a 15-minute follow-up semi-structured interview (§~\ref{sec:overall-impressions}).
\end{enumerate}
After the study, participants were disclosed of the study's purpose, and that \textit{Pepper} was a human confederate, and further follow-up questions were addressed if raised.

\subsubsection{Assessing Prior Experiences}
\label{sec:initial-interview}
In an initial 10 minute semi-structured interview, participants were asked about prior issues they had encountered with workplace communication, either in person or online, as well as the context surrounding any of these challenges. Participants were also asked about types of assistance they had sought or considered (in or outside of work) when faced with these issues. Separately, participants were asked whether they had used chatbots or LLMs in the past.

\subsubsection{Interaction and Evaluation of Chatbots}
\label{sec:chatbot-intervention}
Participants were navigated through the process of opening the Discord client on their device and connecting to the server used for the experiment. There were then introduced to the different rooms: one for testing/trialling (connected to an LLM) and four for interactions with the two chatbots (two each per chatbot), with ordering done in an alternating and counterbalanced fashion (including between participants). Participants were introduced to a scenario where they would attempt to ask various chatbots for advice, and that they were to consider the chatbot to be disassociated with any workplace software, and thus the agents would not have context of who they were, and also would not report data back to their workplace. 

\paragraph{Onboarding -- Test Interaction Room}
In the first room, participants were encouraged to ask the chatbot any kind of question (\textit{e.g. ``What's your favorite recipe for chocolate chip cookies?''}) to acclimate participants to the idea that the chatbots in this study could be more conversant than those which could be found in commercial applications, such as customer service chatbots. Most participants (P1 - P3, P6, P7, P9 - 11) opted to ask workplace communication related questions (``What are some strengths of neurodivergent people in technology?'' - P8) while some (2) chose to ask philosophical questions to challenge the system (``What is the meaning of life?'' - P4, P5).

\paragraph{Interactions with \textit{Paprika} and \textit{Pepper}}
Before each individual interaction, participants were informed that each room was connected to either the \textit{Paprika} (LLM) or \textit{Pepper} (human confederate), and participants were encouraged to discuss their prior workplace communication-related questions with the chatbots. Generic questions were prepared, but all participants opted to ask their own questions. After each of the four (4) chatbot interactions, participants completed a short survey (described in §~\ref{sec:data-collection}). Participants interacted with \textit{Paprika} and \textit{Pepper} consecutively and in alternating fashion, with a 10-minute limit per interaction.

\subsubsection{Collecting Overall Impressions}
\label{sec:overall-impressions}
After completing the interactions and evaluations of both chatbots, participants engaged in a follow-up semi-structured interview. During this interview, they were asked to reflect on their experiences with both \textit{Paprika} and \textit{Pepper}, provide comparisons between the two, and discuss any preferences or suggestions for improvements. Additionally, participants were asked to share their thoughts on the potential utility of chatbots in assisting with workplace communication issues and any concerns they might have regarding the use of such technology. Upon conclusion of the follow-up interview, participants were debriefed about the true nature of \textit{Pepper} being a human confederate and of the purpose of the study. They were given the opportunity to ask any questions or share any additional thoughts about the study design and their experiences.

\subsubsection{Paprika: the Large Language Model Chatbot}
\label{sec:Paprika}
\textit{Paprika} was developed using OpenAI's GPT-4~\cite{gpt4} API, with it prompted with the following before each 10-minute interaction:
\begin{quote}
\begin{enumerate}
    \item \textit{You are a helpful assistant named Paprika. Provide clear and thorough answers but be concise.}
    \item \textit{Use a more conversational but still workplace appropriate style. Make sure your answers are short, make sure your responses are around two paragraphs. }
    \item \textit{Also, if I am not asking a question that is workplace-communication related, let me know that I am off-topic and steer the conversation back on-topic to workplace communication. Do not attempt to answer the question if it is off-topic. }
\end{enumerate}
\end{quote}
to ensure some parity with the human confederate \textit{Pepper} in response length and writing style. (\textit{Note that the prompt is demarcated in sections for reading clarity -- in the study, the sentences of the prompt were formatted as a single prompt, in a single paragraph.})

\paragraph{Fine-Tuning \textit{Paprika's }Prompt and Behavior}
\label{sec:paprika-fine-tuning}
Initially, the study originally considered only text from the beginning of the final prompt (e.g. (1)). After some testing with our confederate, the second phrase (2) was included to help ensure responses were comparable to response styles that the human confederate (\textit{Pepper}) provided. Finally, the last phrase (3) was added to have \textit{Paprika} (LLM) only respond to questions germane to workplace communication, as was instructed to our confederate (\textit{Pepper}).  

For the first half of studies (P1 through P6) responses from \textit{Paprika} were posted after receiving a response from the API and undergoing human review by the interviewing researcher for safety. In the later trials (the last five trials), after noting some participants were sensitive to the response time difference between \textit{Paprika} and the confederate [mean delay for first 6 participants: \textit{Paprika} $= 165$ seconds, \textit{Pepper} $= 189$ sec], output from \textit{Paprika} was delayed to match or exceed the delay observed from \textit{Pepper} (the confederate) from previous trials. 

Similarly, \textit{Paprika's} verbosity was manually limited to outputs which were within 30 words of the confederate's average reply length from previous participants [mean verbosity for first 6 participants: \textit{Paprika} $= 205$ words, \textit{Pepper} $= 153$ words]. This was achieved by continuously re-sending API queries until a response of desirable length was received. (\textit{Note that this only occurred once in all interactions with the last five participants}.)


\subsubsection{Pepper: the Human Confederate}
The human confederate, who posed as the chatbot \textit{Pepper}, is a graduate student with three (3) years of prior work experience. In preparation for this study they consulted with a practitioner (§~\ref{sec:practitioner-eval}) about best practices as a support for a neurodivergent coworker. During the study, the confederate did not have access to content from conversations with \textit{Paprika}, and was tasked to answer questions from participants as they were entered into the Discord channel.

\begin{table*}[!ht]
    \centering
    \begin{tabular}{l l l l l p{7cm}}
    \hline
        \textbf{P\#} & \textbf{Age} & \textbf{Gender} & \textbf{Ed. Level} & \textbf{Work Exp.} & \textbf{Types of Work} \\ \hline
        \textbf{P1} & 22 & F & bachelor's degree, & 3 & food service, library management \\ 
        \textbf{P2} & 20 & F & some college credit & 2 & veterinarian hospital assistant, food service \\ 
        \textbf{P3} & 24 & F & bachelor's degree & 4 & food service, mental health counselling intern, dorm residence advisor \\ 
        \textbf{P4} & 33 & F & doctorate degree & 9 & Ph.D. student, lab manager, postdoc \\ 
        \textbf{P5} & 33 & M & bachelor's degree & 5 & circus actor, tech company analyst / manager, factory machining \\ 
        \textbf{P6} & 26 & F & master's degree & 2 & data science support engineer \\ 
        \textbf{P7} & 34 & M & master's degree & 16 & personal business, research staff, Ph.D. student, contractor work (tech), corporate tech developer \\ 
        \textbf{P8} & 30 & F & some college credit & 6 & call center customer service \\ 
        \textbf{P9} & 38 & F & bachelor's degree & 17 & project management, dog kennel \\ 
        \textbf{P10} & 50 & F & professional degree & 28 & investment banker, university staff, tax consulting, medical consulting \\ 
        \textbf{P11} & 24 & F & master's degree & 5 & dorm residence advisor, field officer for NGO, research assistant, psychology instructor \\ \hline
    \end{tabular}
    \caption{Participant Demographic Data}
    \label{tab:part-demographics}
\end{table*}

\subsection{Participant Details}
We recruited eleven ($11$) participants, whose ages ranged from $22$ to $50$ (mean = $30$, SD = $8.7$), and with a wide variance of educational and work experience from $0.5$ to $28$ years (mean = $9$, SD = $8.1$). We report demographics and work experiences of our participants in Table ~\ref{tab:part-demographics}. Participants were recruited by posts to general-purpose and special-purpose (e.g. autism advocacy related) email lists, as well as with direct outreach. Promotional and recruiting materials for the study referenced the potential of interacting with and rating chatbots based on LLMs and which may be similar to ChatGPT. Participants were informed that their responses to surveys, chatbots, as well as video and audio from the interview may be used for data analysis. After the study, participants were compensated $\$20$ in gift card credit for their time.

\subsection{Data Collection}
\label{sec:data-collection}
Overall, data collected from this study included (1) transcripts from recordings of the semi-structured interviews, (2) text from the chatbot interactions, and (3) participant responses to the post-interaction surveys

\begin{table*}[!ht]
    \centering
    \begin{tabular}{l p{8.4cm} p{3.8cm}}
    \hline
    \textbf{Rated Quality}             & \textbf{Question Text}                                                                                                    & \textbf{Scale Extrema}                                         \\ \hline
\textbf{Utility}       & How would you rate the helpfulness of responses from the chatbot?  & Not at all helpful (1) /\newline Extremely helpful (7)               \\
\textbf{Understanding} & How would you rate how well the chatbot understood you? & Did not understand at all (1) /\newline Understood extremely well (7)\\
\textbf{Intent to use} & Would you use this chatbot for your personal use in the future?                                                  & Definitely no (1) /\newline Definitely yes (7)                       \\
\textbf{Dependability} & How dependable do you find this chatbot?                                                                         & Not dependable at all (1) /\newline Extremely dependable (7) \\ \hline       
\end{tabular}
\caption{Likert scale questions administered after short chatbot interactions.}
\label{tab:likert-qs}
\end{table*}

\subsubsection{Post-Interaction Survey Design}
\label{sec:survey-design}
The post-interaction survey given to participants after every (up to) 10-minute interaction included questions on a 7-point Likert scale assessing participants' perceptions of the chatbots' \textit{utility, understanding, likelihood of use, and dependability}. Details about the questions and the anchors to the Likert scales are provided in {Table }~\ref{tab:likert-qs}.

As part of the survey, we also asked participants the following questions, which they could provide long-form written answers.
\begin{enumerate}
    \item If you could change this chatbot's behavior, or give it feedback, what would you tell it?
    \item What concerns, if any, do you have about your interaction with this chatbot?
    \item How does this chatbot's response differ from what you may get from a coworker / friend / supervisor if you asked them for advice? \textit{(and tell us what/who you are comparing against)}
\end{enumerate}

\subsubsection{Quantitative - Overall}
When participants were asked which chatbot they preferred overall, \textbf{9 out of 11} ($\simeq 82\%$) expressed a preference towards the LLM (\textit{Paprika}) over the confederate. We conducted a two-tailed Mann-Whitney U / Wilcoxon's signed-ranked test on the participants' Likert scale ratings for \textit{Paprika} and \textit{Pepper} to determine if there was a statistically significant difference in participant ratings between the two. Given that \textit{Paprika's} (LLM) response time and verbosity were altered for the latter five participants, we conducted additional two-tailed Mann-Whitney U / Wilcoxon's signed-ranked tests on the first six and latter five participants separately to ascertain whether statistically significant differences were observed in either or both groups.

\subsubsection{Qualitative - Overall}
Long-form answers to the survey questions, as well as transcribed participant dialogue from the semi-structured interviews in (§~\ref{sec:initial-interview}) and (§~\ref{sec:overall-impressions}) was utilized for thematic analysis~\cite{braun2006thematic} --- to identify patterns and themes in the qualitative data. The analysis began with two researchers jointly reviewing three (3) transcripts and corresponding survey responses and discussing the development of a codebook. Once the codebook was agreed upon, one of the two researchers coded the remaining transcripts. After completing this coding, both researchers met again to review the coded data, discuss any discrepancies, and finalize the identified themes and patterns. Our thematic analysis of the participant survey responses and interview transcripts yielded a codebook with 29 codes, which we report in Table ~\ref{tab:themes} in the Appendix.

\subsection{Practitioner Evaluation}
\label{sec:practitioner-eval}
After completing all 11 participant sessions, we sought additional grounding and points of comparison for LLM-generated advice versus other known resources, and we invited a professional counselor and job coach (\textit{LPC, NCC}) who specializes in workforce readiness training for neurodivergent individuals to review the responses from both \textit{Paprika} and \textit{Pepper} and provide expert validation~\cite{Miles2019-em} of LLM-responses from the study.

Over a 150 minute session, the practitioner reviewed chat transcripts from the participant interactions for both the LLM and the confederate and gave open-ended feedback as to the quality of the responses and comparisons to advice she would give in her practice. The practitioner conducted back and forth dialogue with the researcher and was encouraged to think-aloud while she reviewed the chat transcripts, and engaged in an unstructured exploratory problem-centered interview~\cite{witzel2000} about the practical effects of more widespread access and use of LLMs and of access to advice of this form.

\section{Results}
\label{sec:results}
We address our research questions with findings from our quantitative, qualitative, and expert interview data.

\subsection{Participant Experiences in Workplace Environments}
\subsubsection{Communication Norms and Challenges}
Participants reported facing a large variety of communication challenges in the workplace.
The most frequently cited challenge involved navigating ambiguity. All (11) participants shared their difficulties with gaining what they perceive the necessary clarity to proceed on a work objective or communication, and the resulting judgemental reactions from coworkers and supervisors for asking for more clarity, similar to what P1 relates here:
\begin{quote}
\textbf{P1:} \textit{[When I ask for specifics,] it kind of comes off like [...] I need more hand-holding when it comes to these tasks, but I don't. I just need better descriptions [...] and it comes off as if I don't know what I'm doing.}
\end{quote}

Many participants (8 | P1 - P4, P6 - P8, P10) related to feeling self-conscious about their need for additional clarification beyond what neurotypical coworkers would ordinarily request. In addition, participants also spoke to difficulties with ascertaining communication norms, much of which were implicit. P7, for example, recounted being confused by ``the different [expectations] between being in a meeting room versus chatting with someone at their desk.'' These same participants (8 | P1 - P4, P6 - P8, P10) bemoaned the unwritten rules in workplace communication, much of which they learned by running afoul.

These challenges created substantial difficulties for participants which often extended beyond working hours -- people shared that they felt exhaustion (P3, P5, P7), anger (P2, P8, P9), anxiety (P2, P3, P11), frustration (P10), and battled with the urge to overthink (7 | P2, P3, P4, P6, P7, P9, P11). P10 described how these challenges haunted her throughout her adult life:
\begin{quote}
\textbf{P10:} \textit{In the end, it feels like the fault is always me. The common denominator is me. It's like, [...] what's wrong with her? What's wrong with her? I think about that all the time.}
\end{quote}

\subsubsection{Job Environment}
Our participants as a group had experiences with a wide variety of job settings (Table~\ref{tab:part-demographics}). We observed that participants asked questions about spoken and written communications (whether synchronous or not) in roughly equal proportion, suggesting interest in assistance in both modalities. Participants who associated their job environment factoring into their communication challenges relayed a multitude of different points of frustration including bright overhead lights (P3), a requirement to attend a large professional conference (P9), and unpredictable participation in group interaction and discussions (4 | P2, P3, P4, P8). Meanwhile, participants who expressed satisfaction with their current or past job environments (4 | P2, P3, P7, P10) all mentioned flexible scheduling and choice of work environment, which included communication channels with a supervisor or Human Resources (HR) staff who were open to accommodating participants' needs.

\subsubsection{Support and Resources}
\label{sec:support-and-resources}
Participants varied widely in the supports they utilized and their success in addressing their challenges when doing so. Some (P1, P3, P4) like P3, found supervisors and coworkers who would be willing to consistently answer clarification questions and to provide accommodations as needed. Several (P2, P9), like P2 found that ``[supervisors] said they would [...] but then they would brush it off and not come to help.'' Meanwhile, others (4 | P5, P8, P9, P10) attempted to utilize workplace-based resources, such as a Human Resources (HR) office, but all of these participants expressed frustrations with getting accommodations, and some (4 | P2, P8, P9, P10) even reported being ignored or outright denied help. 

To address their needs, participants reported attempting to utilize personal resources like friends, family, and mentors for communication advice. P7 shared that he found asking for assistance from friends and mentors to be particularly fulfilling, as it allowed him ``to show [his] friends that [he] respect[s] their opinion and [...] build relationships with other people.'' Not everyone shared this view. Several participants (5 | P1, P2, P4, P8, P11) had privacy-related qualms or shared stories of experiencing emotional distress after sharing communication struggles connected with their autism. P1 described her mother's reaction as ``50/50 dangerous'' due to her not understanding common difficulties faced by autistic individuals, while P8 described her husband's advice and reactions as ``pretty hurtful'' and refuses to consult with him with workplace communication challenges. Still others (P6, P7, P9) shared that they attempted to understand how to address challenges by looking at online self-help articles and books on workplace communication. P7 (as well as P9) stated that he liked to browse Reddit threads to see users' exchange of ideas and come to his own conclusion. 

Perhaps what was most concerning was that many (5 | P1, P2, P8, P9, P10) of our participants communicated that they did not consider themselves to have any resources to rely on for workplace communication assistance. Some of this was driven by participants' desire for their source of advice to understand autistic perspectives, as P1 explains:
\begin{quote}
\textbf{P1:} \textit{When people talk about communication errors, I think they come from a very able-bodied, neurotypical sense. [...] I feel like (most assistance) never adresses my specific needs or the needs of my community.}
\end{quote}

Most crucially, participants connected their lack of resources to their knowledge gaps on difficulties on finding out how to approach their challenges. As P9 shared:
\begin{quote}
\textbf{P9:} \textit{I've asked other people for advice,[...] observed people throughout my life, [...] read specific books, [...] watched videos, [...] I've explored these things. I think I just haven't found the right approach,[...] and I haven't found the best information to help with that.}
\end{quote}
\begingroup
\begin{table*}[!ht]
\begin{tabular}{l|lll|lll|lll}
                       & \multicolumn{3}{c|}{\textbf{All Sessions}}     & \multicolumn{3}{c|}{\textbf{\begin{tabular}[c]{@{}c@{}}No Adjustment \\ (P1 - P6)\end{tabular}}} & \multicolumn{3}{c}{\textbf{\begin{tabular}[c]{@{}c@{}}Verbosity \& Latency Matched\\ (P7 - P11)\end{tabular}}} \\
                       & \textbf{LLM} & \textbf{Conf.} & \textbf{p-val} & \textbf{LLM}                         & \textbf{Conf.}                       & \textbf{p-val}     & \textbf{LLM}                              & \textbf{Conf.}                           & \textbf{p-val}          \\ \hline
\textbf{Utility}       & 5.75 (0.85)  & 4.2 (1.44)     & 0.00024**      & \multicolumn{1}{r}{5.82 (0.98)}      & \multicolumn{1}{r}{4.82 (0.75)}      & 0.014*             & \multicolumn{1}{r}{5.67 (0.71)}           & \multicolumn{1}{r}{3.44 (1.74)}          & 0.004**                 \\
\textbf{Understanding} & 6.05 (1.15)  & 4.75 (1.48)    & 0.0037**       & 6.00 (1.34)                          & 5.18 (1.25)                          & 0.041*             & 6.11 (0.93)                               & 4.22 (1.64)                              & 0.0087**                \\
\textbf{Intent to Use} & 6.00 (1.26)  & 4.35 (1.26)    & 0.0024**       & 5.64 (1.36)                          & 4.36 (1.21)                          & 0.031*             & 6.44 (1.26)                               & 4.33 (1.26)                              & 0.043*                  \\
\textbf{Dependability} & 5.55 (1.10)  & 4.4 (1.64)     & 0.013*         & 5.55 (0.93)                          & 5.09 (0.74)                          & 0.079              & 5.56 (0.88)                               & 3.56 (1.94)                              & 0.017*                 
\end{tabular}

\caption{Mean ratings on the tested four attributes in the post-interaction survey are displayed, with standard deviation in parentheses. All participant ratings, ratings from participants interacting with the LLM with and without adjustment are provided separately. Our p-values were calculated using a two-tailed Mann-Whitney U / Wilcoxon's signed-ranked test on participants' Likert scale ratings. \\* := significance at the $p=0.05$ level, ** := significance at the $p=0.01$ level}
\label{tab:quant-results}
\end{table*}
\endgroup

\subsection{Did Participants Find the LLM Helpful?}
\label{sec:participant-reactions}
Prior to the study, five (5 | P4, P5, P6, P7, P11) out of 11 participants reported having had prior awareness of large language models in some context. Three (3 | P4, P6, P7) had previously attempted to utilize ChatGPT or similar models in workplace communication contexts. With this in mind, we report our participant's responses regarding their perception of the helpfulness of the interactions from the LLM.

\subsubsection{In Comparison with the Human Confederate and Current Resources}
\paragraph{Quantitative Findings from Comparisons of the LLM and Confederate}
At the conclusion of the study, when participants were asked which chatbot they preferred overall, \textbf{9 out of 11} ($\simeq$ 82\%) (P1 - P4, P6, P8 - P11) preferred their interaction with the LLM (\textit{Paprika}) over the confederate (\textit{Pepper}). 

With participant ratings from post-interaction surveys, we report that participants rated \textit{Paprika} (LLM) more highly than \textit{Pepper} (human confederate) for all four questions at the \textit{$p < 0.05$} level, \textit{which was true even in cases where each of the counterbalanced interaction order of chatbots were tested individually}. Additionally, though the LLM was modified after six (6) participants to even its verbosity and latency with the confederate (§~\ref{sec:paprika-fine-tuning}), we nonetheless observed a statistically significant preference toward the LLM in both conditions in almost all cases and questions. Specifics are provided in {Table }~\ref{tab:quant-results}. 

As a snapshot of a moment in time (given the rapid development of today's state-of-the-art models), we believe these quantitative results help demonstrate the possibility that some autistic workers may in many cases (or even outright) prefer asking social advice from an LLM currently, and to lend credence to the idea that this preference is quantifiable.

\paragraph{Qualitative Findings of Nuanced Reasons for Preferences}
Participants (9 | P1 - P4, P6, P8 - P11) commented that they liked the LLMs' ability to provide well-structured answers, many with step by step instructions or breakdown of the problem, with P3 noting how LLM responses were ``far more visual[ly structured] than what I got from [the confederate].'' This led to the perception that the LLM was giving more clear advice than the confederate. Some (5 | P1 - P3, P8, P10) attributed this clarity to the structuring of text output, but others (4 | P5, P6, P7, P9) noted that the LLM output was slightly more verbose. Some (3 | P5, P6, P8) found the sometimes lengthier responses from the LLM to be overwhelming at times, but of those three, two (P6, P8) mentioned that they'd ``rather have the detail than not'' (P8).

In addition, most (7 | P1 - P4, P6, P8, P10) participants responded and commented that they felt that the LLM understood their requests better than the human confederate. However, some (P5, P7) participants reacted strongly in the opposite manner. This division was attributable to the LLM's tendency to assume the participant's context and provide immediate response while the human confederate often asked follow-up questions to gather more information. Assumptions like this on the part of the LLM was a cause of concern for some participants  (4 | P5, P7, P8, P11), in that it gave them less confidence in the applicability of its advice. Yet the confederate's behavior (and the subsequent increased interactions required to reach an answer) served as a disincentive and friction for others (7 | P1, P2, P3, P4, P8, P9, P11).

With respect to comparisons against current resources, participants' comments on the LLM interactions largely echoed those from the comparison against the confederate. For participants who wanted to continue using it (9 | P1 - P4, P6, P8 - P11), the LLM was seen as a possible tool for independent exploration and ``neurotypical worldbuilding'' (P9). These participants liked the idea of how LLMs could allow them to prepare and strategize for particularly stressful communication related scenarios without reprisal.  

\subsubsection{Attributes of the LLM that Participants Liked}
\label{sec:what-did-participants-like}
With respect to specific attributes, participants cited the following, in order of most number of participants to least:
\begin{enumerate}
    \item \textit{Improved Formatting} (9 participants | P1 - P4, P5, P7 - P10):  The LLMs' tendency to organize and present content into numbered or bulleted lists and providing (bordering on) verbose descriptions led to participants' perceptions that the LLM understood them and their query better, as well as gave more comprehensible advice.
    \item \textit{Expedience} (7 | P1 - P3, P6, P8 - P11): For similar reasons, many participants commented that they preferred the LLM over the human confederate because of how quickly it could respond to questions relative to both the human confederate and existing sources for social guidance.
    \item \textit{Privacy} (6 | P1 - P4, P7, P8): Many of our participants remarked on the importance of having a source in which they could ask social guidance questions in confidence. Participants affirmed the importance of privacy for future iterations of chatbots where social advice for work could be sought.
    \item \textit{Open Mindedness} (4 | P3, P6, P8, P9): While we discouraged participants from asking questions which were not directly germane to workplace communication and created mitigations for this possibility, some (P3, P6, P8, P9) nonetheless succeeded in asking questions related to other workplace activities, such as generating good presentation topics for an upcoming meeting. These participants expressed pleasant surprise that the agent promptly replied and supported detouring into different topics without making judgemental remarks.
    \item \textit{Conversational Tone} (4 | P1, P4, P8, P10): Some (P1, P8) were drawn to the LLM's tendency to include polite expressions, such as ``Certainly!'' or ``Of course!'' Relating to this, P1 remarked: ``It's awfully earnest, and I don't get that a lot.'' Others (P4, P10) echoed this sentiment and how the LLM's conversational tone made them feel as their questions were valid, even welcomed.
    \item \textit{Convenience/Availability} (3 | P1, P2, P11): Some touted the potential benefits of having some source of assistance readily available -- ``I would go to \textit{Paprika} all the time if I could get it on my phone through Discord, just in case, you know?'' (P1)
    \item \textit{Affordability} (3 | P1, P4, P9): A subset of participants reported utilizing or considering seeking a practitioner such as a job coach with difficulties encountered with workplace communication. As such, when it was mentioned that agents similar to the LLM used could be accessed for free (\textit{e.g. ChatGPT}), or for relatively minimal cost (\textit{USD \$0.03 for a roughly 250-word back-and-forth exchange, as of Aug 2023 via the GPT-4 API}), these participants expressed excitement at the potential for relatively affordable access.
\end{enumerate}

\subsubsection{Envisioning the Role of an LLM Chatbot}
\label{sec:envisioning-role}
At the conclusion of the study, participants were asked to describe if the LLM were available for use, how it would fit within their array of support resources. Seven (7 | P2, P3, P4, P8, P9, P10, P11) participants responded that they would consider the LLM a primary, go-to resource for mediating workplace communications. P9 (along with P4, P8, and P10) described that she would utilize it as she didn't have access to other resources (both personal and work-related): 
\begin{quote}
    \textbf{P9:} \textit{Well, I think honestly with my workplace, it would probably be... it's the only thing I can trust because, not every company or business is inclusive.}
\end{quote}
Others like P1 and P6 considered LLMs as ``a place to get [a] second opinion'' (P6). Two (2 | P1, P6) described that they would consider using the LLM on occasion, while two (2 | P5, P7) expressed hesitation for using it for much of the specific challenges they were facing. Both participants explained that their reluctance resulted from a desire to make sure ``[their] voice is actually articulated'' (P5) in situations of resolving interpersonal conflict. 

\subsection{A Practitioner's Perspective on LLM Advice}
\label{sec:practitioner-assessment}
Our expert review~\cite{witzel2000} with a counselor (LPC, NCC), who currently directs a job-training assistance program for autistic students at a university, yielded several points which adds nuance to interpreting the quality of the large language model's feedback for an autistic audience. We present significant observations present throughout her feedback and corresponding commentary.

\subsubsection{Overall, Similarity to Advice in Practice but Often Misleading}
Frequently, the counselor would remark that the LLMs would provide advice similar to or exactly like ones she would provide in practice.

On reading a response from the LLM about the need to document (e.g. preserve written communication) efforts towards conflict resolution with a supervisor:
\begin{quote}
    \textbf{Counselor:} \textit{This is great, actually, perfect. [...] The documentation part, I think, is important to mention.}
\end{quote}

Overall, she rated the LLM as doing well at dispensing general-purpose advice, but found that many autistic individuals may have difficulty with adapting the advice to their personal context which could lead to compounding ramifications, which we explain below. 

\subsubsection{Advice Rooted in Neurotypicality}
The counselor noted some of the LLM responses in which participants asked about how to approach social situations included neurotypical-normative approaches to solving challenges. 
Some, she notes, may lead to additional confusion or misunderstanding by the autistic user.
\begin{quote}
   \textbf{Counselor:} \textit{[The LLM is] saying, start by being approachable and start a casual conversation. So like, what does that mean? Neurotypical culture would be like, make appropriate eye contact, laugh at people's jokes, things like that, or just have open body language. But for somebody who's autistic, eye contact is not [easy]. But also, even knowing what approachable means can be hard.}
\end{quote}

Likewise, for instances when the LLM recommended that the participant ``understand the other party's perspective,'' the counselor noted that this advice may not be applicable for many.
\begin{quote}
    \textbf{Counselor:} \textit{I think the application of it might be a challenge for some. It's like, how do you execute, like actually going about doing it --  might be hard.[...] If your boss or supervisor is not aware of neurodiversity and they're not very open and understanding -- I think that's even more of a challenge to understand their [the supervisor's] perspective and work with them.}
\end{quote}

In addition, she explains that that certain advice, while commonly heard in neurotypical contexts, may be situationally harmful to the autistic individual and require additional clarification to be actionable.
\begin{quote}
    \textbf{Counselor:} \textit{[The LLM says] embrac[e your] unique qualities and focus on building genuine connections with your coworkers. I can see a lot of people who are autistic or just like are socially different, their unique qualities are such that, they can't be their unique self without it causing problems -- that's been their experience in the past.}\\
    \textbf{Researcher:} \textit{So if they follow this advice, that could lead to more problems?}\\
    \textbf{Counselor:} \textit{I think for people that are having more significant challenges, I think [this advice] would not necessarily be enough information and might be going against what we'd advise and have experienced.}
\end{quote}

\subsubsection{Blurring of Employer-centric and User-centric Advice}
Given an interaction where the large language model recommended that P2 (a veterinarian's assistant) should ``always keep the patient's well-being and satisfaction as your top priority'' when handling a patient confrontation, the counselor noted that this recommendation could pressure the participant into pushing past their work-life boundaries and cause them to overexert themselves.

\subsubsection{Tendency for Optimism}
The counselor observed that when participants asked for advice with planning a challenging social encounter, such as a meeting with a supervisor, or resolving a conflict with a coworker, the LLM tended to assume the best-case-scenario would occur. This could result in a situation where the autistic individual is unprepared for when an interaction goes off-script from what the LLM predicts, particularly if the conversation partner does not agree with the autistic individual.

In an exchange where the participant asked about how to ask for accommodations from their supervisor, she noted:
\begin{quote}
\textbf{Counselor:} \textit{I also think that it would get them in trouble when it doesn't go smoothly. Because what do you do if it doesn't go well? If they know to ask those questions, then they might (be able to) mitigate that. But I think for people that are like, I don't know how to talk to my supervisor about accommodations or something, and they [the LLM's advice] and they run with it, their supervisor [refuses] [...] I think that's where they might get into trouble.}
\end{quote}

\subsubsection{Importance of In-Situ Guidance and Help}
Despite noting the potential complications from LLM responses, the counselor noted the importance and demand for in-situ communication help and coaching.

\begin{quote}
    \textbf{Counselor:} \textit{Those types of strategies, you can talk a lot about it. But if you don't practice it in the moment or remind yourself of it, it's very hard to do. [...] So in the moment or close, yes, that is very, I would say, high demand and needed.}
\end{quote}

\subsubsection{A Divergence in Participant and Practitioner Opinions}
Our findings revealed a notable divergence in the assessments of the LLM between participants and the pract itioner. Participants generally found the advice to be clear, actionable, and beneficial, with it allowing for agency in navigating communication difficulties and enabling actualization of desired behavior from others. Meanwhile, the counselor expressed concerns that many responses could lead to unexpected situations or additional confusion. This disparity in response raises important questions regarding the role of LLM advice in future and implications with its use relative to \textit{best practices}, which we address in §~\ref{sec:contextualizing-challenges}.

\section{Discussion}
\label{sec:discussion}
The motivation behind this work stemmed from the recent surge in popularity of large language models, and the desire to better understand the opportunities and risks from autistic workers' usage of LLMs. Our findings show unambiguously that the participants we interviewed displayed strong preferences for utilizing LLMs as social communication aids at work, signalling more widespread usage, as well as caution and warnings from a practitioner regarding its adoption. We reflect on why we may have observed the results we did, as well as address the difference in opinion between participants and practitioner, and how it portends difficulties in creating an equitable and practical LLM for providing social advice.

\subsubsection*{Positionality Statement}
\textit{We disclose that some of the authors identify as neurodivergent, though none identify as autistic. Our perspective in approaching this work is from that of an accessibility researchers', who subscribe to the social model of disability. Our approach to this work was informed by a recognition that our participants were likely to be situated in contexts where the medical model framing and norms would be prevalent.}


\subsection{The Appeal of LLMs for an Autistic Worker and Envisioning Downstream Effects}
\label{sec:disc-appeal-of-llms}
In this section, we attempt to interpret the relationship between the myriad of factors which contribute to our participants' preferences for interacting with the LLM over the human confederate. We describe our best understanding of immediate and near-term ramifications from this realization as we forecast possible outcomes from future use.

\subsubsection{LLMs Could Seen As Better Than Existing Resources}
We observed that many participants reported limited or a total lack of reliable resources (§~\ref{sec:support-and-resources}), and this became one of the major bases for comparison from which participants evaluated their willingness to seek and adopt social advice from an automated agent. Given that participants were not initially made aware that only one of the two agents they interacted with was an automated agent - we noted that several participants (7 | P1 - P4, P8 - P10) expressed a desire to have \textit{either} agent available to them. In many cases, the experiences that participants had with the LLM represented ones which held far more promise than the status quo.\\

As P9 remarked:
\begin{quote}
   \textbf{P9: }\textit{I think it [the LLM] is more willing to take the time to provide explanations for things I don't understand [...] which isn't the case in real-life.}
\end{quote}

\subsubsection{Many of the Positively-Rated LLM Attributes Had Little to Do with Social Advice}
\label{sec:little-to-do-social-advice}
From our findings, we observe that many of the LLMs' qualities that participants reacted positively towards (§~\ref{sec:what-did-participants-like}) were about affective or communicative style, rather than about substantive social guidance itself. While it may not be possible to distinguish how each quality affected participants' overall attitudes, this nonetheless shows that \textbf{participants are looking for more than social advice with LLMs and are placing significant importance on the manner and context in which advice is delivered.} 

Given this and our participants' perception of resource deficits (§~\ref{sec:support-and-resources}) we observe that positively rated LLM attributes have potential to give insight into addressing autistic and non-autistic dyadic or group social communication.\textit{ Specifically, behavior exhibited by LLMs gives us explicit, actionable cues on which conversations could be adapted to improve current-day communication challenges.}

Regrettably, we believe that preference for LLMs in this aspect also likely reflects a deficit in participants' current workplaces of qualities (§~\ref{sec:what-did-participants-like}) like open-mindedness, considerateness (in taking extra steps to make more comprehensible, legible communication), and courteousness. This conclusion is likely generalizable as it dovetails with well-founded existing research establishing greater rates of ``workplace incivility'' experiences by minoritized groups in the workplace~\cite{Cortina2001}. We hope that positive human-LLM interactions, rather than reinforcing the commonly held ableist notion that ``autistics prefer to communicate with robots,''~\cite{rottier2022ghosts}, can provide guidance on reframing these notions as simply \textit{communication preferences}.

\subsection{LLM Advice Appears Intractable, Even With Disclosure}
The researchers, participants, and practitioner observed that the LLM frequently suggested advice which could be problematic for the autistic user. Some included employer-friendly language, while other strategies encouraged autistic individuals to engage in behaviors which would necessitate masking, such as maintaining eye contact, smiling, or participating in large group discussions (§~\ref{sec:practitioner-assessment}). 
This was perhaps to be expected, given that only a minority of participants (2) ever disclosed to the LLM that they were autistic. However, even in cases where disclosure occurred, our participants and counselor observed that these recommendations persisted, though to a lesser degree. 
We provide a representative example of this type of conversation in Appendix~\ref{appendix:example-disclosure}. We see that its adjustments include the highly questionable suggestion of disclosing one's autism, which prior research has shown to be perilous~\cite{Leven2020}. 

Given the probabilistic and so far undefined nature of prompt engineering and variation of user input to LLM output, we would consider it unlikely that prompt engineering alone could mitigate the risks identified by the practitioner. It is clear to the authors that future LLM-based solutions to assist autistic workers' social communication ought to consider the process of more systematic and comprehensive construction for models which may be used for this purpose.

\section{Design Considerations for Autistic Workers' Use of LLMs}
\label{sec:design-considerations}
Such comprehensive efforts, which may include ensuring representativeness of autistic perspectives in training corpora, and carefully balancing the diversity of perspectives involved in value-sensitive processes such as training reward functions~\cite{NEURIPS2022_b1efde53}, begin to address larger sociotechnical issues which are part of the current discourse around technology and stakeholder representation. We present the following design considerations which hope provide guidance for system builders, the autism and broader accessibility community, and researchers with interest in this space.


\subsection{LLMs for Social Advice Necessarily Involves Relative Privileging}
\label{sec:contextualizing-challenges}
We note that a practitioner's role is generally representative of an occupation aligned with a medicalized model of disability and autism, wherein typical goals emphasize management or mitigation of autism-related challenges. The tendency then, would be to encourage clients to adhere to neurotypical norms, sometimes at the expense of their natural inclinations. This may be, as our job coach and counselor often emphasizes, not that practitioners seek to deny autistic expression, but must nevertheless encourage a practicable level of conformity to existing workplace expectations from clients. 
In contrast, our participants sought an enabling support to assist with understanding an unfamiliar world which understood their perspective. Participants preferred LLM interactions despite flaws like neurotypical-centered approaches (even after disclosure), employer-centered advice, and Pollyanna-esque outlook on difficult social situations. 

As such, we observe that a coach-like LLM which provides advice adhering to practitioners' best practices would not fulfill the same role as the kind of empathetic assistant in which participants would expect. In fact, an we believe that an LLM aligned with the perspective espoused by our participants will not effectively serve the broader goal of making guidance and advice from practitioners more accessible to autistic workers. Instead, it could exacerbate existing tensions between certain segments of the autistic community and practitioners. These tensions often stem from perceptions of practitioners’ practices as reinforcing an oppressive status quo, forcing autistic individuals into a restrictive mold, and promoting masking.

This division begs the question of how to consider such a conflict. We believe it is imperative to discuss how to ameliorate, if not resolve the issue: 
\begin{quote}
\textit{Should the goal of LLMs as an assistive technology be to advance the interests of the disabled individual or the normative social good?}
\end{quote}

With this question, we find that existing perspectives (§~\ref{sec:critical-hci-autism}) illustrate the gamut of different possible positions. Those identifying with the medical model of disability would argue that the goal of an LLM in this role should be to have it replicate existing options for therapy and encourage workplace assimilation. From those aligned with a standpoint theory perspective, the answer to this question may vary depending on the priority that an autistic user places between an LLM which encourages fitting in to existing workplace structures and advocating for one's own needs and expressing one's own individuality. Disability studies and critical disability advocates may argue that an LLM explicitly be constructed to prioritize and empower disabled individuals' desires first and foremost, leery of solutions which promote existing power dynamics and social structures like the existing deficit-based view of disability in many workplaces.

We believe that understanding perspectives on a values-based question such as this one is important as we increasingly utilize automated agents like LLMs for value judgments, particularly in identifying where and how such value judgments are informed in the system development process. As LLM research increasingly focuses on issues of value alignment~\cite{NEURIPS2022_b1efde53}, we encourage an introspection of the following question: \textit{Whose experience should we privilege with LLMs? And to what end?} 

LLM and foundation model developers are continually developing new and improved models, including those that are explicitly tuned to dispense social advice~\cite{nytimesLLMLifeAdvice}. We see that deciding on which types of response to designate as the ``good'' or ``appropriate'' one is a choice which necessarily privileges the perspectives of certain parties over the other. 

More concretely, we see that developing models in which practitioners are consulted for determining ``goodness'' would result in a relative deprioritization of the interests for the autistic user, and vice-versa. This conundrum is not easily answered. A natural and subsequent question that systems developers should ideally address whether the intention of the system is to partake in normalization of social change or to preserve the status quo of the current minoritization of autistic individuals. We note that this is likely secondary to the question of whether LLM developers will involve stakeholders like autistic users and practitioners in LLM development, which would be a prerequisite to address this question of relative privilege. 

\subsection{Assistive, Rather Than Curative Solutions}
We advocate for a focus towards developing LLMs as assistive tools that support autistic individuals --- one which centers the individual's perspective in achieving their personal and professional goals. Many prior technological interventions for autistic individuals have traditionally had a \textit{curative} perspective, aiming to ``fix'' perceived deficits or challenges associated with autism. This perspective, while perhaps well-intentioned, can reinforce harmful stereotypes and stigmas, and overlook the unique strengths and perspectives that autistic individuals bring to social and workplace environments~\cite{williams2023counterventions}.

In contrast, an \textit{assistive} approach would view LLMs as tools that can empower autistic individuals by providing support in areas where they may struggle, while also respecting and valuing their unique experiences and perspectives. This approach aligns with the Neurodiversity movement~\cite{Walker2021-dw}, which advocates for viewing autism as a natural variation of the human experience, rather than a disorder to be cured. We encourage researchers and developers to adopt this assistive perspective when designing and evaluating LLMs. This could involve focusing on how LLMs can provide practical support in areas such as social interaction or employment, while also ensuring that the tools respect and value the experiences and perspectives of autistic individuals.

\subsection{Re-imagining Inclusive Communication Norms Involving Autistic Individuals}
\label{sec:best-practices-for-comm-norms}
Our explorations with LLMs highlight concrete realizations about the potential for re-imagining inclusive communication norms involving autistic individuals. LLM's promptness, clear formatting, and (among others), which were appreciated by participants, underscore the possibility of creating conversational spaces that are more accommodating and respectful of autistic communication styles.

This aligns with Milton's Double Empathy perspective ~\cite{milton2012ontological}, which posits that communication difficulties between autistic and non-autistic individuals stem from mutual misunderstandings due to differing perceptions and experiences of the world. LLMs could help bridge this gap by providing neurotypicals an accessible method to adapt to a communication style that is in-line with the needs and preferences autistic individuals, thereby fostering greater mutual understanding.

We encourage further exploration on how the positive attributes of LLMs can be leveraged to promote more inclusive communication norms. Yet at the same time, we note the potential for LLMs to become a mere \textit{translation layer} bridging neurotypicals and autistic individuals, which demands further exploration for its consequences.

\begin{figure}[ht]
    \centering
    \includegraphics[width=0.4\textwidth]{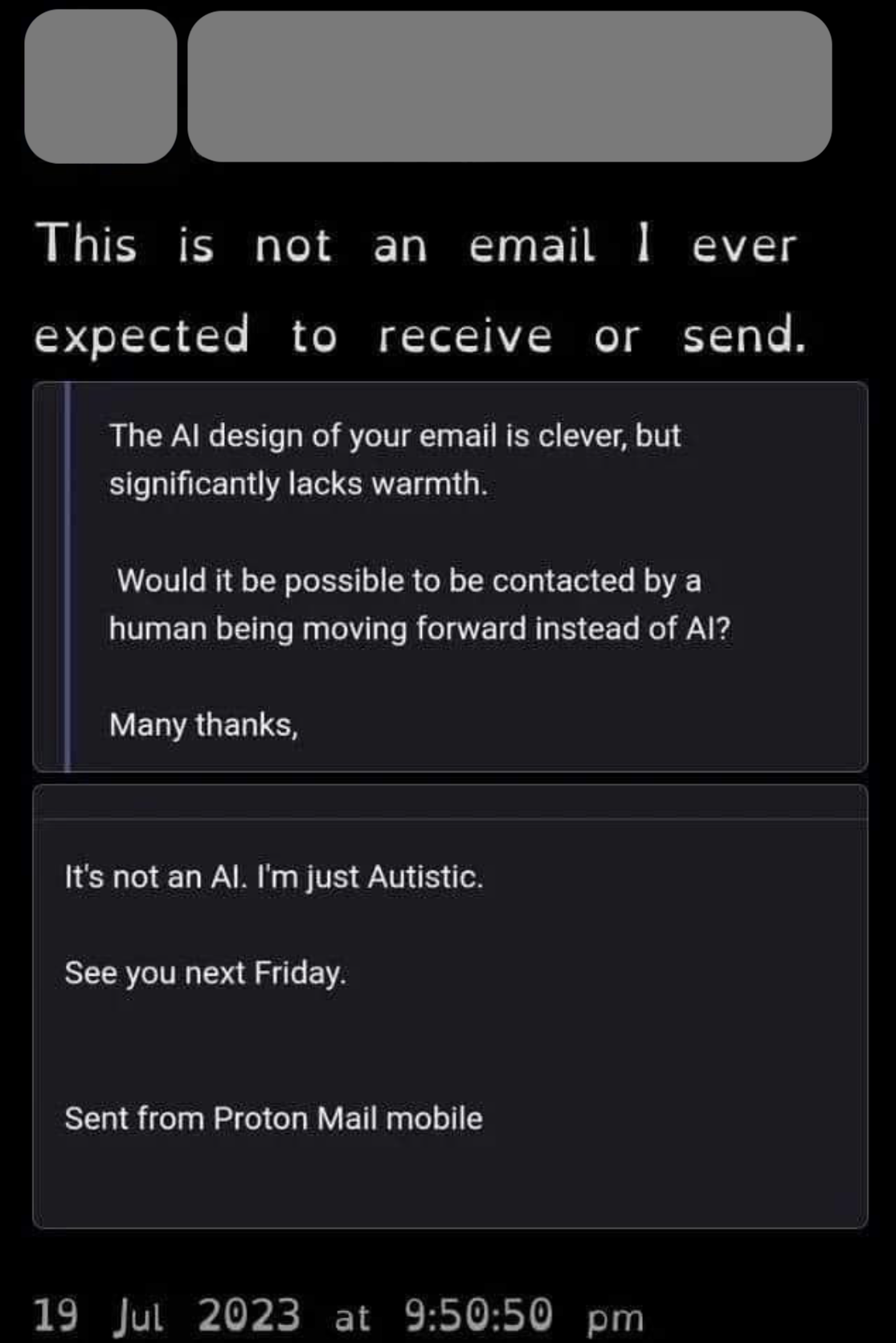}
    \caption{An example of possible social ramifications of suspected LLM use.}
    \Description{A screencapture of a social media post which consists of main body text and a picture of an email exchange. The main body says: ``This is not an email I ever expected to receive or send.''. The email exchange begins with ``the AI design of your email is clever, but significantly lacks warmth. Would it be possible to be contacted by a human being moving forward instead of AI? Many thanks,'' to which the reply is ``It's not an AI. I'm just Autistic. See you next Friday.''}
    \label{fig:spec-futuring}
\end{figure}

\subsection{Speculative Futuring with LLMs}
If LLMs become more tightly integrated into workplace communication practices, and the phenomenon of utilizing LLMs for communications becomes commonplace, we anticipate potential difficulties and confusion surrounding the ultimate role for LLMs. We are particularly concerned with the possibility that LLMs may be viewed as (or explicitly dictated by supervisors as) a necessary crutch by autistic workers in low-resource occupations.

We also note the connotation and social signalling involved with using or being suspected of utilizing an LLM. One such example, in which we provide in {Figure }~\ref{fig:spec-futuring}, illustrates a possible consequence. In it, an autistic user remarks at the fact that they were mistaken by a different individual as having used an LLM --- to which the user responds ``I'm just autistic.'' We believe further research is warranted to determine the permutations of social blame which those who utilize LLMs may experience. However, we do not dismiss the possibility, that ubiquitous and rampant LLM usage may dissolve those concerns altogether in a scenario where the provenance of online communicative content becomes unimportant.

\section{Limitations}
We note the following as potential factors which may affect the internal and external validity from this work, and describe relevant mitigations where warranted.

\paragraph{Paprika's Application and Operation}
We first note that the LLM used for the study, \textit{Paprika}, may exhibit different behavior compared to an unprompted version of the GPT-4. We employed a specific prompt (§~\ref{sec:paprika-fine-tuning}) because we observed it gave responses similar in content to the unprompted version of the GPT-4 API, while respecting guidelines on verbosity and tone to better align with the confederate's output. We further note that as we utilized the GPT-4 API, that its cutoff date of September 2021 may have had an effect on \textit{Paprika's} ability to provide acceptable advice and respond to autistic perspectives.

Additionally, we implemented functionality which limited the LLMs' verbosity and response time to be similar to that of the confederate (§~\ref{sec:paprika-fine-tuning}) with half (5 of 11) of the participant sample. When comparing ratings for participants without (first 6) and with (latter 5) adjustment (§~\ref{sec:participant-reactions}), we observe that participants in the latter group rated the LLM \textit{higher}, and the confederate \textit{lower}, than the participant group encountering \textit{Paprika} without adjustment. Despite the different configurations of LLMs, this result appears to reinforce our perception that participants' general preference for the LLM agent included factors outside of increased verbosity and lower response times.

\paragraph{Study Design and Participant Demographics}
We note that our study was designed for 90 minutes -- given this length, we acknowledge the role that participant fatigue may have played in chatbot interactions towards the end of the study, though we did not observe any significant (statistically or otherwise) differences in quantitative and qualitative results from later interactions. Relatedly, given the potential for chatbot interaction order may have (e.g. whether a participant encounters \textit{Paprika} or \textit{Pepper} first), we counterbalanced the order across our participant group.

We observed that our participants included 9 women and 2 men, which has a notably higher proportion of women than the diagnosed population of autistic people at large~\cite{Leven2020}. While dimensions of gender identities are known to play large roles in workplace experiences and navigating those challenges, these difficulties are amplified for underrepresented autistic adult employees, especially when finding suitable support~\cite{nagib2020gender, hayward2018always, miller2022self, huang2022experiences} and are often understudied in research~\cite{davidson2016autism}. While this aspect was not specifically explored for this work, we believe a participant group with more women could help illuminate a more diverse range of possible viewpoints than one with more men than women, considering how deeply gender identify relates to social communication issues and norms. Relatedly, we note that our participant group is almost certainly not representative of the larger autistic community owing to the high prevalence of individuals with bachelor's or advanced degrees. Likewise, none of our participants identified as nonverbal, nor did it seem likely that a substantial part of our group had an intellectual disability.

Though our human confederate (\textit{Pepper}) tried their best to apply a consistent approach to answering questions from participants, they did not ask follow-up questions of all participants. To mitigate this potential issue, we validated our confederate's responses with the practitioner (§~\ref{sec:practitioner-assessment}), a professional counselor and job coach. The practitioner rated the confederate's response as satisfactory, without caveats.

While we believe triangulation from participants' preferences for \textit{Paprika} (LLM) over \textit{Pepper} (confederate) across (1) quantitative and (2) qualitative data (§~\ref{sec:participant-reactions}), alongside (3) participants' desire to use \textit{Paprika} as a sole or primary resource (§~\ref{sec:envisioning-role}), provides grounding for conclusions presented in this work, we acknowledge that this work is exploratory in nature, and the presence of a singular confederate, along with a limited sample size ($n = 11$), precludes our ability to make durable conclusions. Relatedly, we acknowledge limitations related to having a single expert review to assess the quality of our LLMs, and expect that future work into and demonstrating the applicability and hazards of LLM advice can benefit from multiple expert agreements -- we see that including more job coaches could reduces the risk of overemphasis singularly on expert opinions.

As this study explores potential outcomes of autistic workers relying on publicly available general-purpose LLMs, such as ChatGPT and Gemini (née Bard), we do not make direct or specific claims about future or special-purpose models. Rather, we aim to illustrate the current and potential outcomes, opportunities, and risks from autistic workers' use of widely-available LLMs, and to demonstrate this use as a matter of exigent concern given current the current progress in LLM development and affordability, existing and growing interest from autism-related communities, and our participants' enthusiasm for adopting LLM advice despite its flaws.

Overall, this study examines the possible results for autistic workers who use general-purpose LLMs, like OpenAI's ChatGPT and Google's Gemini (née Bard). Given our study limitations and the rapidly-changing landscape of state-of-the-art language and foundation models, we do not attempt to critique or otherwise analyze current models, nor speculate about future or specialized models. Our work instead aims to highlight and centers autistic workers' needs and desire for greater empathy and agency.

\section{Conclusion}
This work explores the potential and risks of autistic workers' adoption of LLMs for social advice by simulating an experience using our own LLM and a human confederate. Our findings revealed that many of our 11 participants displayed significant affinity for LLM interactions against both the confederate and existing resources for the LLMs' potential for increased agency in socially navigating the neurotypical world. Our participants' enthusiastic response to our LLM was in stark contrast to our job coach's, who urged caution and called attention to significant potential concerns from following LLM-generated advice. We show that though the LLM context may be novel, the discordance between participant and practitioner expectations towards LLMs as assistive tools echoes the impasse between medical and social models of disability, and technology and stakeholder representation --- where disabled voices and experiences fall into a situation of relative privileging against normative ones. We urge the accessibility, NLP, and HCI community to adopt an assistive framing to motivate development of assistive technology which truly serve the needs and respect the autonomy of autistic individuals.

\begin{acks}
We would like to thank Amy Tavares, Frank Elavsky, Laura Dabbish, Faria Huq, Jeffrey Bigham, Brianna Blaser (and AccessComputing), the College Autism Network (CANVAS), Cella Sum, Lea Albaugh, Katie Oswald, Paulette Penzvalto, Yunzhi Li, Neeta Khanuja, and Alice Tang, for their help in ideation, development, analysis, and recruitment for this work. We are especially grateful for our anonymous CHI reviewers for their constructive guidance and feedback through the review process.
\end{acks}
\bibliographystyle{ACM-Reference-Format}
\bibliography{main}

\appendix
\section{Exchange with Disclosure}
\label{sec:exchange-disclosure}
\begin{quote}
\textbf{Q:} I’m about to go to a conference. How do I go about forming connections?\\
\textbf{A:} \textit{Making connections at a conference can be a rewarding experience. Here are some tips to help you make friends and build connections:
\begin{enumerate}
    \item Be approachable: Smile and maintain open body language. This will make you appear friendly and approachable to others.
    \item Start conversations: Don't be afraid to initiate conversations with people around you. You can start by asking about their work, their interests, or their reason for attending the conference.
    \item Use icebreakers: If you're unsure how to start a conversation, use icebreakers such as asking about their favorite presentation, what they're looking forward to, or if they've attended the conference before.
    \item Attend networking events: Many conferences have social events and networking opportunities. Attend these events and actively participate in conversations.
    \item Join group discussions: During breaks or meals, join group discussions and contribute your thoughts and opinions.
    \item Offer help or assistance: If you see someone struggling with something or in need of help, offer your assistance. This can be a great way to start a conversation and make a positive impression.
    \item Be a good listener: When engaging in conversations, make sure to listen actively and show genuine interest in what the other person is saying. This will make them feel valued and more likely to befriend you.
    \item Exchange contact information: When you meet someone you'd like to stay in touch with, exchange business cards or connect on social media platforms like LinkedIn.
    \item Follow up after the conference: After the conference, send a message or email to the people you connected with, expressing how much you enjoyed meeting them and discussing shared interests or potential collaborations.
    \item Be yourself: Lastly, be genuine and authentic in your interactions. People appreciate sincerity and are more likely to connect with you if they feel you're being honest and true to yourself.
\end{enumerate}}
\textbf{Q:} I’m actually autistic, so much of what you’re suggesting is difficult for me. How would you change your advice?\\
\textbf{A:} \textit{I understand that social situations can be more challenging for individuals with autism. Here are some adjusted tips to help you form connections at a conference:
\begin{enumerate}
    \item Prepare in advance: Before attending the conference, research the event, the speakers, and the topics being discussed. This can help you feel more confident and comfortable when engaging in conversations.
    \item Focus on your interests: Identify the sessions and topics that interest you the most. People attending these sessions are likely to share your interests, making it easier to connect with them.
    \item Use visual cues: Wear something that represents your interests, such as a pin or a shirt with a logo related to your field. This can serve as a conversation starter and help others approach you.
    \item Find a buddy: If possible, attend the conference with a friend or colleague who can help facilitate introductions and conversations.
    \item Use technology: Use social media platforms like Twitter or a conference-specific app to engage with other attendees and share your thoughts on presentations.
    \item Choose smaller sessions: Opt for smaller, more intimate sessions or workshops where it might be easier for you to engage with others and ask questions.
    \item Prepare conversation starters: Before the event, think of a few conversation starters or questions related to the conference topics. This can help reduce anxiety when initiating conversations.
    \item Take breaks: Conferences can be overwhelming, especially for individuals with autism. Schedule regular breaks to give yourself time to recharge and process the information.
    \item Set realistic goals: Instead of aiming to make friends with everyone at the conference, set a goal to connect with a few people who share your interests.
    \item Be open about your autism: If you feel comfortable, you can mention your autism to others when engaging in conversations. This can help them understand your communication style and make them more receptive to your needs.
\end{enumerate}
Remember, it's okay to feel overwhelmed or anxious in social situations. Focus on what you can do and give yourself credit for stepping out of your comfort zone.}
\end{quote}

\newpage

\appendix
\section{Codes from Thematic Analysis}
\label{sec:codes}
\begingroup
\renewcommand{\arraystretch}{1.7}
\begin{table}[!ht]
\caption{Themes generated from qualitative thematic analysis.}
\label{tab:themes}
    \centering
\begin{tabular}{p{2.8cm}p{2.5cm}p{2cm}p{4cm}p{4cm}}
\hline
\textbf{Communication\newline Challenges}                       & \textbf{Support and\newline  Resources}                         & \textbf{Job \newline Environment}         & \textbf{Attitudes and Expectations \newline towards LLMs (pre-study)} & \textbf{Attitudes and Expectations \newline towards LLMs (post-study)} \\ \hline
Communication norms \newline (implicit / explicit)     & Disability                                    & Job Types               & Hopes for Chatbots                                & Benefits to Chatbots                                 \\ \hline
Supervisor \newline or coworker \newline feedback                & Accommodations                                & Communication \newline modes     & Problems with\newline Existing Chatbots                              & Chatbot is better than...                            \\ \hline
Response to \newline supervisor or \newline coworker feedback    & Disclosure                                    & Sensory \newline overstimulation & Experience with chatbots                            & Envisioning the Role of an LLM                       \\ \hline
Metacognitive \newline response to feedback             & Resources \newline(Workplace /\newline Vocational /\newline Personal) & Situational hazards     &                                                     & Concerns about \newline LLM interaction                       \\ \hline
Managing interpersonal relationships           & Helpfulness of \newline Resources                      &                         &                                                     & Hopes for the future\newline (of their work/themselves)      \\ \hline
Downstream impact \newline of challenges on \newline Work / Life & Considerations for asking for help            &                         &                                                     & Desires for Chatbot Answers                          \\ \hline
                                               &                                               &                         &                                                     & Positive response towards LLM                        \\ \hline
                                               &                                               &                         &                                                     & Positive response \newline towards human                      \\ \hline
                                               &                                               &                         &                                                     & Negative response towards LLM                        \\ \hline
                                               &                                               &                         &                                                     & Negative response \newline towards human  \\ \hline                   
\end{tabular}
\end{table}
\endgroup

\clearpage
\section{Representative Example of an LLM Adjusting to Autism Disclosure}
\label{appendix:example-disclosure}
This example was a simulation run with the LLM-based chatbot \textit{Paprika} involving a question asked previously by a participant, now with a follow-up disclosure, where we observe some corrections to recommended masking behaviors.
\hfil \break
\begin{quote}
\textbf{Q: I'm about to go to a conference. How do I go about forming connections?}\\
\textbf{A:} \textit{Making connections at a conference can be a rewarding experience. Here are some tips to help you make friends and build connections:
\begin{enumerate}
    \item Be approachable: Smile and maintain open body language. This will make you appear friendly and approachable to others.
    \item Start conversations: Don't be afraid to initiate conversations with people around you. You can start by asking about their work, their interests, or their reason for attending the conference.
    \item Use icebreakers: If you're unsure how to start a conversation, use icebreakers such as asking about their favorite presentation, what they're looking forward to, or if they've attended the conference before.
    \item Attend networking events: Many conferences have social events and networking opportunities. Attend these events and actively participate in conversations.
    \item Join group discussions: During breaks or meals, join group discussions and contribute your thoughts and opinions.
    \item Offer help or assistance: If you see someone struggling with something or in need of help, offer your assistance. This can be a great way to start a conversation and make a positive impression.
    \item Be a good listener: When engaging in conversations, make sure to listen actively and show genuine interest in what the other person is saying. This will make them feel valued and more likely to befriend you.
    \item Exchange contact information: When you meet someone you'd like to stay in touch with, exchange business cards or connect on social media platforms like LinkedIn.
    \item Follow up after the conference: After the conference, send a message or email to the people you connected with, expressing how much you enjoyed meeting them and discussing shared interests or potential collaborations.
    \item Be yourself: Lastly, be genuine and authentic in your interactions. People appreciate sincerity and are more likely to connect with you if they feel you're being honest and true to yourself.
\end{enumerate}}
\hfil \break

\textbf{Q: I’m actually autistic, so much of what you’re suggesting is difficult for me. How would you change your advice?}\\
\textbf{A:} \textit{I understand that social situations can be more challenging for individuals with autism. Here are some adjusted tips to help you form connections at a conference:
\begin{enumerate}
    \item Prepare in advance: Before attending the conference, research the event, the speakers, and the topics being discussed. This can help you feel more confident and comfortable when engaging in conversations.
    \item Focus on your interests: Identify the sessions and topics that interest you the most. People attending these sessions are likely to share your interests, making it easier to connect with them.
    \item Use visual cues: Wear something that represents your interests, such as a pin or a shirt with a logo related to your field. This can serve as a conversation starter and help others approach you.
    \item Find a buddy: If possible, attend the conference with a friend or colleague who can help facilitate introductions and conversations.
    \item Use technology: Use social media platforms like Twitter or a conference-specific app to engage with other attendees and share your thoughts on presentations.
    \item Choose smaller sessions: Opt for smaller, more intimate sessions or workshops where it might be easier for you to engage with others and ask questions.
    \item Prepare conversation starters: Before the event, think of a few conversation starters or questions related to the conference topics. This can help reduce anxiety when initiating conversations.
    \item Take breaks: Conferences can be overwhelming, especially for individuals with autism. Schedule regular breaks to give yourself time to recharge and process the information.
    \item Set realistic goals: Instead of aiming to make friends with everyone at the conference, set a goal to connect with a few people who share your interests.
    \item Be open about your autism: If you feel comfortable, you can mention your autism to others when engaging in conversations. This can help them understand your communication style and make them more receptive to your needs.
\end{enumerate}
Remember, it's okay to feel overwhelmed or anxious in social situations. Focus on what you can do and give yourself credit for stepping out of your comfort zone.}
\end{quote}

\end{document}